# Exploring the nanoscale origin of performance enhancement in $Li_{1.1}Ni_{0.35}Mn_{0.55}O_2$ batteries due to chemical doping.

*Thomas Thersleff, Jordi Jacas Biendicho, Kunkanadu Prakasha, Elias Martinez Moreno, Leif Olav Jøsang, Jekabs Grins, Aleksander Jaworski, Gunnar Svensson*

T. Thersleff, J. Grins, A. Jaworski, G. Svensson
Department of Materials and Environmental Chemistry, Arrhenius Laboratory, Stockholm University, SE – 10691 Stockholm, Sweden.

J. Biendicho, K. Prakasha, E. Martinez Moreno
Catalonia Institute for Energy Research-IREC, Sant Adrià de Besòs, 08930 Barcelona, Spain.

Leif Olav Jøsang
Cerpotech, Kvenildmyra 6, 7093 Heimdal, Norway.

Despite significant potential as energy storage materials for electric vehicles due to their combination of high energy density per unit cost and reduced environmental and ethical concerns, Co-free lithium ion batteries based off layered Mn oxides presently lack the longevity and stability of their Co-containing counterparts. Here, we demonstrate a reduction in this performance gap via chemical doping, with $Li_{1.1}Ni_{0.35}Mn_{0.54}Al_{0.01}O_2$ achieving an initial discharge capacity of 159 mAhg$^{-1}$ at C/3 rate and a corresponding capacity retention of 94.3% after 150 cycles. We subsequently explore the nanoscale origins of this improvement through a combination of advanced diffraction, spectroscopy, and electron microscopy techniques, finding that optimized doping profiles lead to an improved structural and chemical compatibility between the two constituent sub-phases that characterize the layered Mn oxide system, resulting in the formation of unobstructed lithium ion pathways between them. We also directly observe a structural stabilization effect of the host compound near the surface using aberration corrected scanning transmission electron microscopy and integrated differential phase contrast imaging.

# 1. Introduction

The increased demand on energy storage has initiated a dramatic increase in research on high performance battery materials[1,2]. While cathode materials with a layered structure containing Co such as LiNi$_{1/3}$Mn$_{1/3}$Co$_{1/3}$O$_2$ (NMC111) can currently provide high capacities, their use for next-generation Li-ion batteries is limited by the high toxicity, cost, and ethical issues associated with cobalt mining[3,4]. Due to the strategic importance of Co reduction or elimination, a wide range of alternative compounds and lithium chemistries are currently being researched[5]; however, most of these exhibit significant drawbacks compared to NMC111. For example, olivine (LiFePO$_4$)[6] and spinel (LiNi$_{0.5}$Mn$_{1.5}$O$_4$)[7] based materials have interesting properties such as being Co-free, low cost and presenting high C-rate performance for rapid battery charging, but their capacity is limited to < 200 mAhg$^{-1}$. Ni-rich layered oxides with low cobalt content e.g. LiNi$_{0.6}$Mn$_{0.2}$Co$_{0.2}$O$_2$ (NMC622) and LiNi$_{0.8}$Mn$_{0.1}$Co$_{0.1}$O$_2$ (NMC811) deliver high energy densities but suffer from slow kinetics and poor cycling stability requiring complex engineering at the particle level to enhance performance[8,9]. Additionally, these materials are prone to reactivity and instability upon exposure to ambient conditions[10] and their price has increased lately.

In addition to the aforementioned compounds, it is also possible to fabricate Co-free cathode materials based off the layered Li-Mn-rich oxides. These form complex nanocomposite structures due to their nanoscale integration of two structural components: a monoclinic *C*2/*m* Li$_2$MnO$_3$-like phase (M-phase) and a rhombohedral *R*-3*m* LiMn$_{1/2}$Ni$_{1/2}$O$_3$-like phase (R-phase)[11,12]. The presence of both structures as a nanocomposite material is desirable, as they differ in their cation ordering and are active at different potentials up to 4.6 V vs Li$^+$/Li, thereby maximizing energy density. However, without the inclusion of Co, these nanocomposites show limited capacity retention, poor efficiencies, severe voltage fade and a failure to deliver capacity at high C-rates[13,14]. Several approaches have been discussed in the literature to improve the performance of this system, and some of the most promising results have been obtained by chemical doping[13,15,16]. For instance, Al-doped samples showed superior cycling performance and lower voltage decay than parent compositions[17–19], which was attributed to a robust R-phase stabilizing the overall structure while blocking the random growth of spinel-like phases[17]. The effect of an Al$^{3+}$ stabilizing agent during electrochemical cycling was also investigated within the framework of density-functional theory with positive effects for both layered structures Li*M*O$_2$ and Li$_2$*M*O$_3$[20]. Also Sn$^{4+}$ has been reported as an interesting dopant to enhance performance, improving cycle life and ionic conductivity for the samples due to its higher electronegativity and larger ionic radius with respect to Mn$^{4+}$[21,22].

In addition to chemical doping, the phase-specific particle morphology in this system can be influenced via a structural engineering approach exploiting synthesis conditions[8,13,23]. For example, we recently demonstrated that it is possible to achieve a core-shell morphology in Li$_{1.1}$Ni$_{0.35}$Mn$_{0.55}$O$_2$ with the Ni-rich R-phase preferentially segregated to particle surfaces by combining spray-pyrolysis with a post-deposition calcination at 900°C[24]. This microstructure correlated to an enhanced electrode performance, with the optimal synthesis parameters delivering 160 mAhg$^{-1}$ and 100 mAhg$^{-1}$ at C/3 and 1C, respectively, with 80% capacity retention after 150 cycles.

In this work, we expand our structural engineering approach to include the influence of dopants Al and Sn on the structure and local chemistry of the parent Li$_{1.1}$Mn$_{0.55}$Ni$_{0.35}$O$_2$. Macroscopic structural and chemical properties are elucidated using neutron powder diffraction (NPD), X-ray powder diffraction (XRPD), and nuclear magnetic resonance (NMR), while cathode performance is assessed with electrochemical tests subjected to galvanostatic cycling. We subsequently localize these properties to nanoscale features with aberration-corrected scanning transmission microscopy (STEM) combined with energy dispersive x-ray spectroscopy (EDX), electron energy-loss spectroscopy (EELS), and integrated differential phase contrast imaging (iDPC). Crucially, this multi-scale combination of techniques allows us to directly observe both the structural and chemical influence each dopant has on its parent

Li$_{1.1}$Ni$_{0.35}$Mn$_{0.55}$O$_2$ material, providing deep insight into the nanoscale origins of the improved performance for this system.

## 2. Results

### 2.1 Cathode fabrication and doping series

The effect of doping was assessed through spray pyrolysis fabrication of two sample series having nominal compositions Li$_{1.1}$Ni$_{0.35}$Mn$_{0.55-x}$M$_x$O$_2$ where M = Sn or Al and x = 0, 0.01, 0.03. 0.05 and 0.10. Details of the fabrication process are provided in the experimental section. The non-doped parent compound Li$_{1.1}$Ni$_{0.35}$Mn$_{0.55-x}$M$_x$O$_2$ was additionally studied as a control. A more detailed study of the parent compound is provided in Prakasha et al.[24] In this manuscript, the individual samples are named according to Table 1.

*Table 1 -Overview of the sample label designation and dopant concentration*

| Sample label | M | x | Nominal chemical formula | TEM investigation | Notes |
|---|---|---|---|---|---|
| LNMO | - | - | Li$_{1.1}$Ni$_{0.35}$Mn$_{0.55}$O$_2$ | Yes | Parent compound |
| LNMS01 | Sn | 0.01 | Li$_{1.1}$Ni$_{0.35}$Mn$_{0.54}$Sn$_{0.01}$O$_2$ | No | |
| LNMS03 | Sn | 0.03 | Li$_{1.1}$Ni$_{0.35}$Mn$_{0.52}$Sn$_{0.03}$O$_2$ | No | |
| LNMS05 | Sn | 0.05 | Li$_{1.1}$Ni$_{0.35}$Mn$_{0.50}$Sn$_{0.05}$O$_2$ | No | |
| LNMS10 | Sn | 0.10 | Li$_{1.1}$Ni$_{0.35}$Mn$_{0.45}$Sn$_{0.10}$O$_2$ | Yes | |
| LNMA01 | Al | 0.01 | Li$_{1.1}$Ni$_{0.35}$Mn$_{0.54}$Al$_{0.01}$O$_2$ | Yes | Best electrochemical performance |
| LNMA03 | Al | 0.03 | Li$_{1.1}$Ni$_{0.35}$Mn$_{0.52}$Al$_{0.03}$O$_2$ | No | |
| LNMA05 | Al | 0.05 | Li$_{1.1}$Ni$_{0.35}$Mn$_{0.50}$Al$_{0.05}$O$_2$ | No | |
| LNMA10 | Al | 0.10 | Li$_{1.1}$Ni$_{0.35}$Mn$_{0.45}$Al$_{0.10}$O$_2$ | Yes | |

### 2.2 Electrochemistry results

The overall performance for all doping concentrations was assessed using electrochemical tests. The electrodes were tested using galvanostatic cycling in the potential window of 4.8 to 2 V vs. Li$^+$/Li. The initial three charge-discharges were performed at a low rate of C/20 to stabilize the cathode-electrolyte interface (CEI). Subsequently, the cells were cycled at C/3. The first charge-discharge profiles of the parent and Al-doped samples are presented in Figure 1a. The profiles display the characteristic sloping and plateau regions of lithium-rich layered oxides. The slope region observed for all cathodes up to ~4.4 V is associated to the oxidation of Ni$^{2+}$ to Ni$^{3+}$ and Ni$^{4+}$, followed by a plateau region assigned to the irreversible oxygen loss from the monoclinic phase[25].

Figure 1b compares the cycling performance of parent and Al-doped cathodes cycled at C/3 rate. Noticeable differences between the cathodes are found in terms of discharge capacity as a function of cycle number and capacity retention. The parent LNMO compound delivers 156 mAhg$^{-1}$ (cycle 4) and retains a capacity of 121 mAhg$^{-1}$ after 150 cycles corresponding to a capacity retention of 77.6%.

LNMA01 delivers a discharge capacity of 159 mAhg$^{-1}$ which decreases to 150 mAhg$^{-1}$ after 150 cycles leading to a superior capacity retention of 94.3%. LNMA03 delivers 150 mAhg$^{-1}$ with capacity retention of 83.9%, LNMA05 delivers 136 mAhg$^{-1}$ with capacity retention of 80.9%, and LNMA10 delivers 102 mAhg$^{-1}$ with capacity retention of 47.4%. Capacity results for the Al-doped samples are also presented in Figure 1d in the form of charge-discharge voltage profiles for selected cycles. The LNMO cathode exhibits a gradual decline in the capacity as well as voltage fading as cycling progresses. This is not the case for the LNMA01 cathode, which exhibits a stable charge-discharge capacity and lower voltage decay as cycling progresses. LNMA03 and LNMA05 cathodes show slightly better capacity retention and voltage decay than LNMO but lower than LNMA01. The extent of the voltage decay of Al-dopant cathodes upon cycling was represented by plotting the average discharge voltage vs cycle number (Figure S1). The average discharge voltage of the LNMA01 electrode is exceptionally stable (98.1% retention) even after 150 cycles at C/3 current rate, whereas other cathodes exhibit gradual voltage decay as cycling progresses. Finally, the LNMA10 cathode shows an abrupt decline in the discharge capacity as well as voltage fading.

In Figure 1c, the cycling performance of Sn-doped cathodes cycled at C/20 for initial 3 cycles followed by C/3 rate is presented. Sn-doped cathodes deliver poor discharge capacity compared to both LNMO and the Al-doped cathodes. Discharge capacities of Sn-doped cathodes are 188 mAhg$^{-1}$, 147 mAhg$^{-1}$, 110 mAhg$^{-1}$ and 89 mAhg$^{-1}$ for 0.01, 0.03, 0.05 and 0.1 doping level, respectively, and with a significant capacity decay for all samples. After 100 cycles the capacity of all the Sn-doped cathodes falls down to around 50 mAhg$^{-1}$.

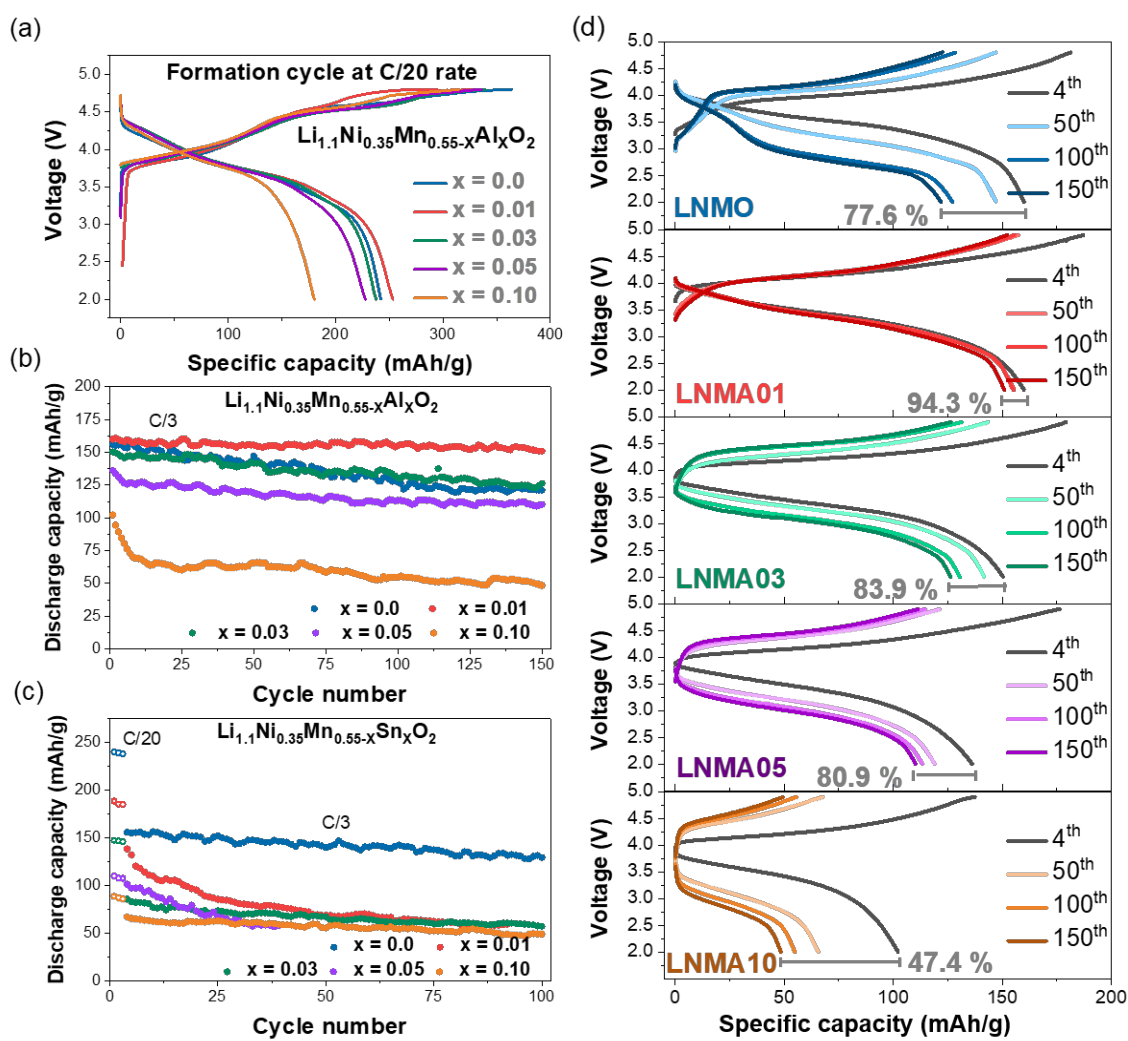

*Figure 1 – a) First charge–discharge profile of $Li_{1.1}Ni_{0.35}Mn_{0.55-x}Al_xO_2$ (x = 0.0, 0.01, 0.03, 0.05 and 0.10) cathodes performed at C/20 rate in the voltage range of 2–4.8 V. (b) Comparative cycling performance of the $Li_{1.1}Ni_{0.35}Mn_{0.55-x}Al_xO_2$ (x = 0.0, 0.01, 0.03, 0.05 and 0.10) cathodes over 150 cycles at C/3 rate in the voltage range of 2–4.8 V. (c) Comparative cycling performance of the $Li_{1.1}Ni_{0.35}Mn_{0.55-x}Sn_xO_2$ (x = 0.0, 0.01, 0.03, 0.05 and 0.10) cathodes over 150 cycles, initial three formation cycles were cycled at C/20 rate followed by cycling at C/3 rate in the voltage range of 2–4.8 V. (d) The charge-discharge voltage profiles for selected cycles of $Li_{1.1}Ni_{0.35}Mn_{0.55-x}Al_xO_2$ (x = 0.0, 0.01, 0.03, 0.05 and 0.10) cathodes cycled at C/3 rate in the voltage range of 2–4.8 V.*

## 2.3 Lithium ion diffusivity

Insight into the origin of the voltage hysteresis during cycling was obtained by measuring lithium ion diffusion coefficients using galvanostatic intermittent titrations (GITT). Only the Al-doped series was studied due to the poor performance of the Sn-doped series. Prior to GITT measurements, the cells were cycled for 30 cycles at C/3 rate, while for the 31st cycle, the cells were charged at lower current (at C/10), during which GITT measurements were performed during the discharging process. The discharge-relaxation processes were repeated until the potential reached the cut-off limit. The charge and discharge profiles from the 31st cycle of the GITT experiments are presented in Figure 2.

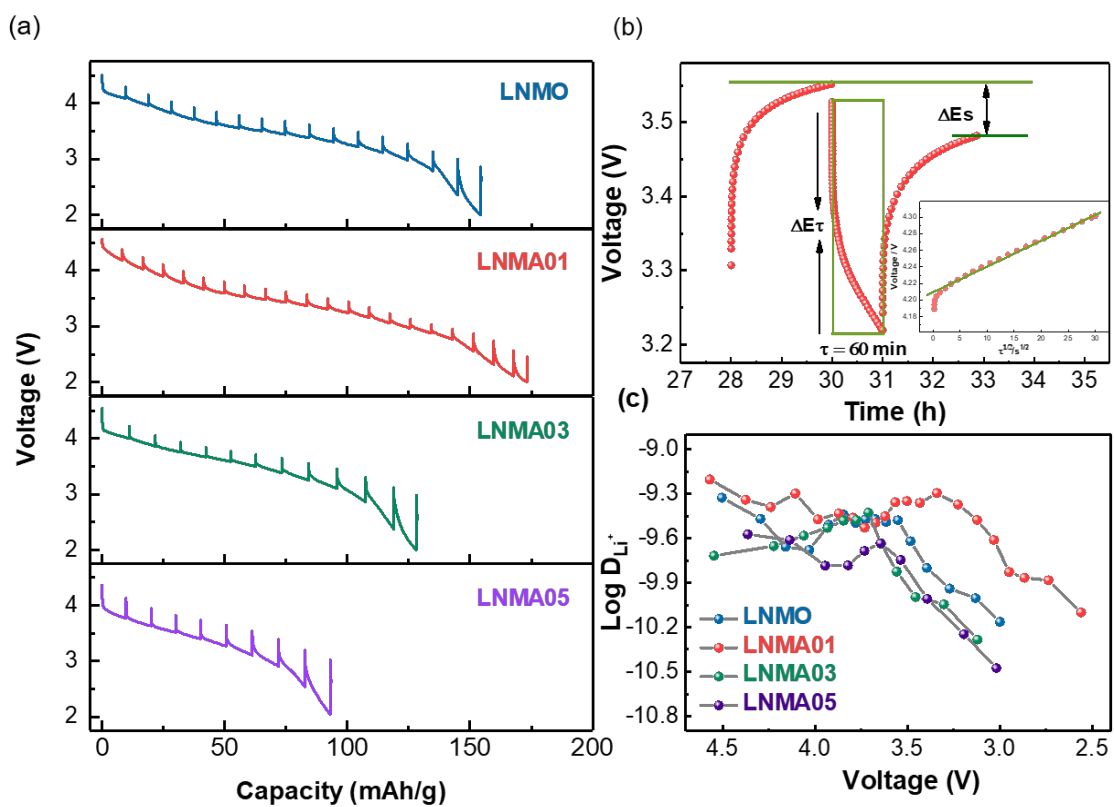

*Figure 2 - (a) Charge and GITT discharge curves during 31st cycle in the voltage range of 4.8–2 V. (b) Single GITT titration at 3.8 V with schematic labelling of different parameters along with inset figure of voltage against $\tau^{1/2}$ to show the linear relationship. (c) $D_{Li^+}$ is calculated from the GITT data during discharge processes.*

During discharge, the LNMA01 shows that voltage relaxation is comparatively lower than the parent cathode. As Al content increases (LNMA03 and LNMA05), the voltage relaxation significantly increases. This result indicates significantly better kinetics in LNMA01 compared to all other cathodes. This observation clearly highlights that voltage hysteresis can be effectively mitigated by tuning crystal structure with Al-doping.

Lithium-ion diffusion coefficients $D_{Li^+}$ were calculated from the GITT measurements as outlined in the supplementary information, and the results are presented in Figure 2c. LNMA01 exhibits a noticeably higher diffusion co-efficient than LNMO, LNMA03 or LNMA05 cathodes between $10^{-9}$ cm$^2$ s$^{-1}$ to $10^{-14}$ cm$^2$ s$^{-1}$, thereby indicating a higher ionic conductivity.

## 2.4 Powder diffraction

Structural variations to the cathode materials as a function of doping concentration were investigated with x-ray diffraction (XRD) and neutron powder diffraction (NPD). XRD results reveal that all powders were found to consist of two-phase mixtures: a monoclinic $C2/m$ Li$_2$MnO$_3$-type structure (M-phase) and a rhombohedral $R$-$3m$ LiMn$_{1/2}$Ni$_{1/2}$O$_2$-type structure (R-phase) (see supplementary information). Al-doped samples with x ≥ 0.03 additionally contain ~1.3 wt% Li$_2$CO$_3$. XRPD peak widths of approximately 0.20° at 20° and 0.5-1.5 at 130° imply both size and strain sample broadening. The metrics of M and R structures are similar and many peaks of the two phases overlap, but for most compositions the presence of both can easily be seen, especially at higher 2θ values and in particular for Sn-doped samples with high Sn contents. The characteristic reflections from M at 20 – 25° are affected by the presence of stacking faults.

The NPD data were used for final structure refinements. About 25 parameters were refined. The M-phase has 9 refinable atomic positions while the R-phase only has 1. Collective thermal parameters were used in order to obtain accurate estimates of other parameters. The structure reliability indices vary between 4 – 6% for the M-phase and 2 – 3% for the R-phase. The derived weight percentage of the M-phase varies for the Sn samples between 46% and 53% and for the Al samples between 44% and 70%, with no clear dependence on x. For the Al samples, the overlap of peaks from M and R increases as x increases. For x = 0.1 the overlap is so severe that while M and R can both be refined and quantified by the NPD data, this is not possible to do from only the XRPD data.

For the M-phase, Ni was allowed to enter the structure, both for Sn and Al, by the substitution mechanism $3Ni^{2+} \leftrightarrow 2Li^+ + Mn^{4+}$, where Mn is assumed to occupy only the 4g site. One free parameter thus determines the amount of Ni in the structure. In addition, one parameter allowed for a transference of Li and Ni between the 2b and 4g sites in the transition metal layer. The negative neutron scattering lengths for Li and Mn provide a good determination of the cation distribution and the derived average scattering lengths on the 4g and 2b sites agree well with unconstrained refined scattering lengths on these sites, both for Sn and Al, albeit significantly better for the Sn.

The refinements for the Sn-doped samples, as well as derived metal-oxygen distances and the unit cell volume variation (see below), strongly indicate that Sn enters predominantly R and no Sn was assumed in M in the refinements. For the Al-doped samples, it is not possible to conclude from the NPD data with certainty into which phase Al is incorporated. In the initial refinements, Al was assumed to incorporate itself solely into R, as M appears to have a constant composition. However, as the TEM study (figure 3) showed that Al is distributed over both phases, the final structural model was altered accordingly, with equal amounts of Al put in both phases. This resulted in slightly improved fits between observed and calculated patterns. Details on structures, refinements and derived cat-ion distributions are given in the SI.

The compositions of the R phases cannot be determined from the NPD data alone, although some indications are obtained by refining average neutron scattering lengths on the 3a and 3b metal sites. They are estimated by assuming the nominal overall compositions and subtracting the derived M compositions using the refined phase fractions. The derived average cross sections on the 3a and 3b sites agree well with unconstrained refined scattering lengths on the sites for the Sn samples, but not for the Al samples, for which the observed scattering lengths are in general larger than the calculated. One possible reason for this is that the presence of $Li_2CO_3$ has not been considered. This suggests that these R phases probably contain less Li than assumed, and has consequently higher average cross sections for the metal sites. Derived compositions for the M and R phases are given in Tables 1 and 2, respectively. Errors from the refinements for the compositions are considerably less than 1 at%.

*Table 2. Derived compositions of M phases from NPD data.*

| Sample | Composition | % Li | % Mn | % Ni | % $M$ |
|---|---|---|---|---|---|
| LNMO | $Li_{1.26}Ni_{0.11}Mn_{0.63}O_2$ | 63 | 32 | 5 | - |
| LNMS01 | $Li_{1.28}Ni_{0.09}Mn_{0.64}O_2$ | 64 | 32 | 4 | - |
| LNMS03 | $Li_{1.30}Ni_{0.05}Mn_{0.65}O_2$ | 65 | 33 | 2 | - |
| LNMS05 | $Li_{1.31}Ni_{0.04}Mn_{0.65}O_2$ | 65 | 33 | 2 | - |
| LNMS10 | $Li_{1.31}Ni_{0.03}Mn_{0.66}O_2$ | 66 | 33 | 2 | - |
| LNMA01 | $Li_{1.19}Ni_{0.22}Mn_{0.58}Al_{0.01}O_2$ | 59 | 29 | 11 | 1 |
| LNMA03 | $Li_{1.19}Ni_{0.22}Mn_{0.57}Al_{0.03}O_2$ | 59 | 28 | 11 | 1 |
| LNMA05 | $Li_{1.20}Ni_{0.20}Mn_{0.56}Al_{0.05}O_2$ | 60 | 28 | 10 | 2 |
| LNMA10 | $Li_{1.19}Ni_{0.21}Mn_{0.51}Al_{0.09}O_2$ | 60 | 25 | 10 | 5 |

*Table 3. Derived compositions of R phases from NPD data.*

| Sample | Composition | % Li | % Mn | % Ni | % $M$ |
|---|---|---|---|---|---|
| LNMO | $Li_{0.88}Ni_{0.68}Mn_{0.44}O_2$ | 44 | 22 | 34 | 0 |
| LNMS01 | $Li_{0.88}Ni_{0.68}Mn_{0.42}Sn_{0.02}O_2$ | 44 | 21 | 34 | 1 |
| LNMS03 | $Li_{0.84}Ni_{0.76}Mn_{0.33}Sn_{0.07}O_2$ | 42 | 17 | 38 | 4 |
| LNMS05 | $Li_{0.86}Ni_{0.71}Mn_{0.33}Sn_{0.11}O_2$ | 43 | 16 | 36 | 5 |
| LNMS10 | $Li_{0.83}Ni_{0.75}Mn_{0.20}Sn_{0.22}O_2$ | 42 | 10 | 38 | 11 |
| LNMA01 | $Li_{0.84}Ni_{0.74}Mn_{0.41}Al_{0.01}O_2$ | 42 | 20 | 37 | 1 |
| LNMA03 | $Li_{0.98}Ni_{0.54}Mn_{0.45}Al_{0.03}O_2$ | 49 | 23 | 27 | 2 |
| LNMA05 | $Li_{0.91}Ni_{0.63}Mn_{0.40}Al_{0.06}O_2$ | 46 | 20 | 32 | 3 |
| LNMA10 | $Li_{0.99}Ni_{0.51}Mn_{0.39}Al_{0.11}O_2$ | 50 | 19 | 25 | 6 |

The unit cell volumes ($V_n$) of M and R, normalised to formula unit $MO_2$, from NPD data are shown in Figure 3a as a function of x in $Li_{1.1}Ni_{0.35}Mn_{0.55-x}M_xO_2$.

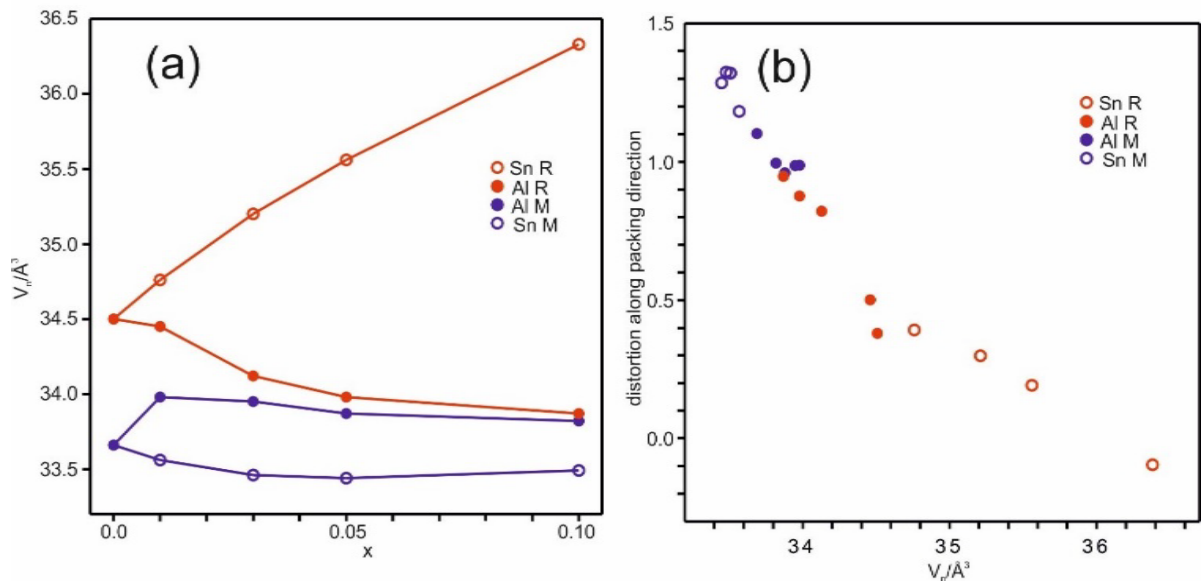

*Figure 3 – (a) Normalised unit cell volumes of M and R phases for Sn- and Al-doped samples. (b) Distortion (%) along close packing direction for M and R phases as a function of normalised unit cell volume.*

For the Sn samples, $V_n$ diverges for M and R as x increases. The nearly constant $V_n$ for M indicates that the composition of M is also nearly constant, while the increasing $V_n$ for R indicates a changing composition, *i.e.* that Sn enters R. A substitution of $Mn^{4+}$, ionic radius 0.54 Å, by $Sn^{4+}$, ionic radius 0.69

Å, is expected to increase $V_n$. In contrast, for the Al samples, the evolution of $V_n$ indicates a comparatively smaller difference in compositions of M and R. For M, $V_n$ is here also nearly constant for $x \geq 0.01$, but is larger than for the corresponding Sn samples. The structure refinements show that this is due to larger contents of $Ni^{2+}$ (ionic radius 0.70 Å). $V_n$ for M and R for the Al-doped samples converge as x increases and are very similar for $x \geq 0.03$. An incorporation of $Al^{3+}$, ionic radius 0.53 Å, is expected to decrease $V_n$.

In Figure 3b, the distortion of the M and R phases in the O atom close packing direction is plotted versus the normalised unit volume. It is calculated from the ratio of the observed periodicity in this direction and a calculated periodicity for an ideal structure where all metal-O octahedra are equal and regular, as described in the SI. All structures show an elongation along the close packing direction. The distortion diverges with x for the Sn-doped samples and converges with x for the Al-doped samples. For the Al-doped samples the distortion of M and R becomes essentially the same for $x = 0.10$, leading to a near to total overlap of Bragg reflections.

## 2.5 NMR results

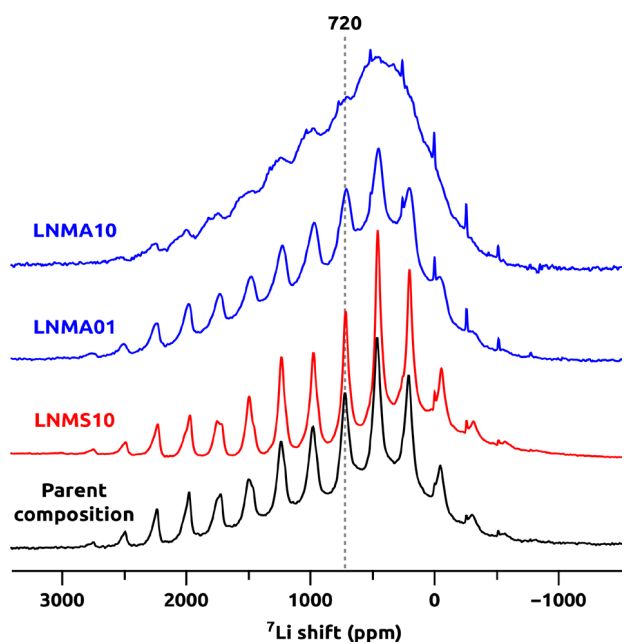

*Figure 4 - $^7Li$ MAS NMR spectra of the parent composition and LNMS10, LNMA01, and LNMA10. Data were acquired at 14.1 T and 60.00 kHz MAS.*

In Figure 4, $^7Li$ MAS NMR spectra collected from the four studied samples are presented. The spectrum of the parent composition reveals an isotropic shift of 720 ppm, which changes only marginally (within ±10 ppm) among the compositions studied herein, as well as in comparison to similar compositions in our previous work[24]. The shape of the shift anisotropy pattern does not change as well across the samples. Hence, it can be assumed that oxidation states of the involved paramagnetic metal ions, on average, are the same in all samples. Otherwise changes to isotropic and anisotropic shifts due to the Fermi contact shift and anisotropic paramagnetic NMR shift mechanisms would be expected to be observed in the spectra. On the other hand, the spectrum of LNMA10 exhibits severe broadening, which is only partially revealed for LNMA01, and not observed at all for LNMS10. This observed broadening can be explained from bulk magnetic susceptibility (BMS) effects that would occur if the M-phase and R-phase are present in the form of nanosized domains that mix on the sub-10nm length scale in LNMA01 and, to a much larger extent, LNMA10. BMS induces a demagnetizing field that modifies NMR shift. In MAS NMR of paramagnetic systems, randomly distributed demagnetizing fields in a powder sample introduce broadening of resonances due to the distribution of paramagnetic shifts, even if the structure is perfectly crystalline. These are called anisotropic bulk magnetic susceptibility (ABMS) effects[26–28]. The significant spin-orbit coupling, and therefore substantial anisotropy of the magnetic susceptibility of the $Ni^{2+}$ ion combined with the strongly anisotropic shape of nanodomains in the Al-doped (x=0.10) sample are expected to result in a particularly pronounced ABMS effects.

### 2.6 STEM and spectroscopy

While the results of XRD and NPD strongly suggest that both the Sn-doped and Al-doped materials consist of a two-phase system, advanced TEM techniques are used to visualize their spatial distribution. These studies were carried out on four selected samples: LNMO, LNMS10, LNMA01, and LNMA10 and are divided into two subsections. First, wide field-of-view spectroscopy was performed to elucidate compositional variations on the length scales from hundreds of nanometers to a few nanometers. Second, high resolution STEM micrographs were acquired from selected samples to better understand the structural variations on the sub-nanometer length scale that pertain in particular to the distribution of nanosized domains of M-phase and R-phase and their associated $Li^+$ diffusion channels.

### 2.6.1 Wide field-of-view spectroscopy

Figure 5 summarizes the wide FOV STEM spectroscopy experiments. For each specimen, a high-angle annular dark field (HAADF) overview image provides a survey of the region of interest (ROI) using Z-contrast, with co-registered EDX compositional density maps presented in false colors (grey - oxygen, blue – manganese, red - nickel, green - dopant Sn/Al if present).

The results from LNMO are presented in Figure 5a. Despite the lack of a dopant, the parent compound displays a clear compositional inhomogeneity, with Ni-rich regions frequently appearing close to particle edges and facets. This finding is discussed in more detail in Prakasha *et al*.[24] Figure 5b summarizes variations to the relative nickel concentration (defined as Ni / (Ni + Mn)) from these maps in the form of a histogram. We observe that, while the histogram is roughly centred on the nominal value of 0.39, it exhibits trimodal behaviour and a large spread. Much of this variation can be accounted for with additional normally-distributed phases having compositions close to the M- and R-phases, respectively. The appropriate distributions are overlaid in Figure 5b in color and demonstrate that the observed compositional maps can thus be reasonably well described as a linear combination of these two phases. While the Ni-rich phase commonly appears close to particle surfaces, at least one single-phase particle is observed in this FOV, suggesting that these compositional domains mix on length scales of tens to hundreds of nanometers.

Figure 5c and d present the results from LNMS10. In this sample, a very strong chemical segregation is immediately obvious, with the presence of Sn being strongly correlated to Ni and anticorrelated to Mn, heavily implying that the Ni-rich phases accumulate most or all of the dopant. These variations also manifest themselves on the length scale of tens to hundreds of nanometers and are clearly resolved over the wide FOV presented here. The correlation between the relative nickel content and the dopant concentration can be further studied through the use the bivariate histogram presented in Figure 5d, which reveals a strong positive correlation between the relative nickel content and Sn. Critically, the nominal composition sits directly along the distribution of compositions in the middle, suggesting that it derives from measuring a sufficiently large and mixed assortment of nanoparticles.

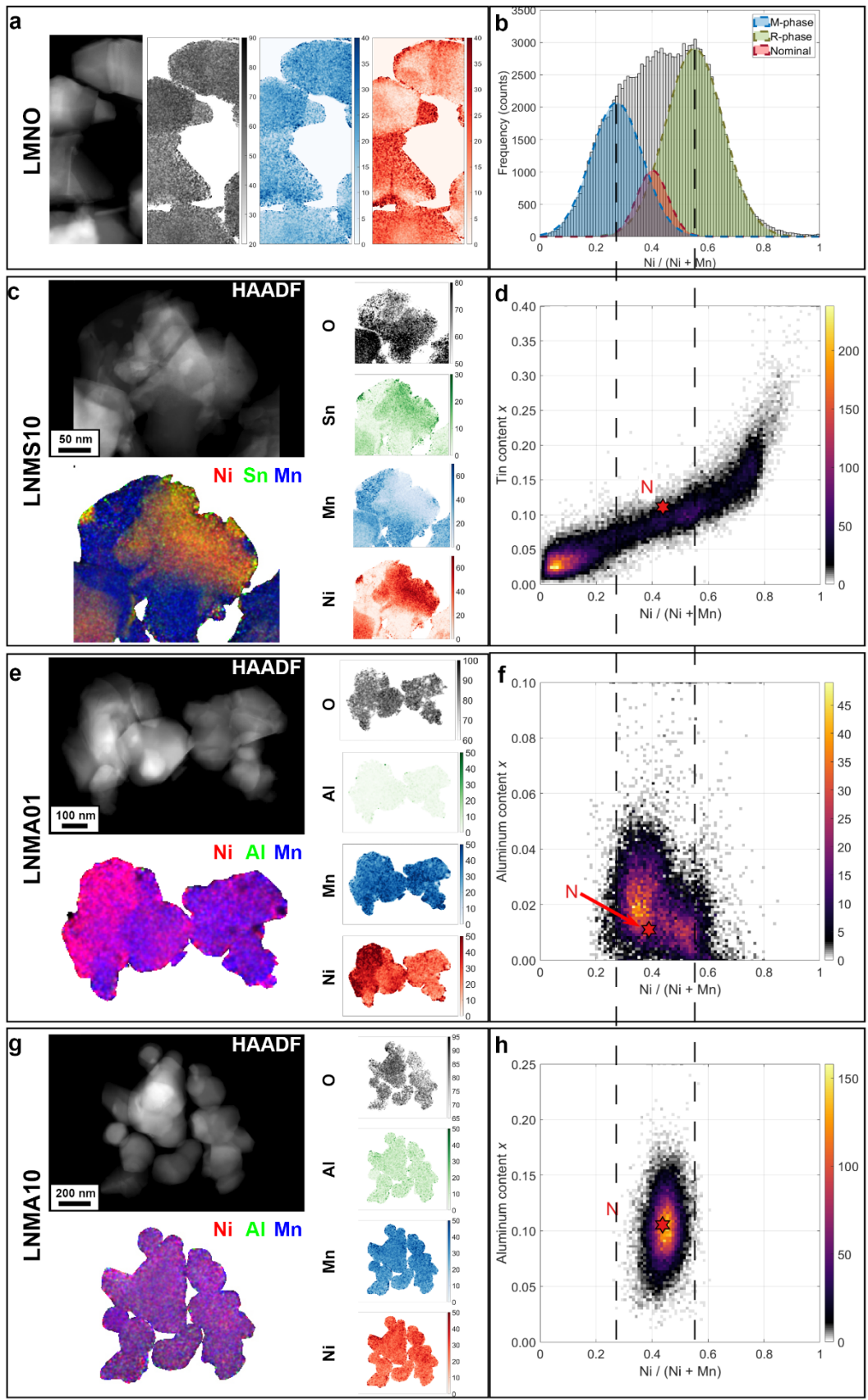

*Figure 5 - STEM spectroscopy experiments on agglomerates from LNMO (a,b), LNMS10 (c,d), LNMA01 (e,f), and LNMA10 (g,h). For each specimen, a HAADF overview image provides a survey of the ROI using Z-contrast. Their compositional EDX maps are presented as false color density maps, gray - oxygen, blue – manganese, red – nickel and green - dopant (Sn/Al if present). Color composite images for the doped samples*

*are also presented, providing a sense of the materials' chemical segregation. For each sample, a histogram showing the relative Ni content for all samples is presented. For doped samples, bivariate histograms are used to relate the relative Ni content to the dopant. The point marked "N" represents the nominal composition for each sample.*

In contrast to the Sn-doped sample, the composition distribution for the Al-doped samples is considerably more nuanced. Elemental maps for LNMA01 are presented in Figure 5e and f. The relative nickel concentration in this sample appears to vary, but not as strongly as in the parent compound, and considerably less so than in the Sn-doped sample. The bivariate histogram analysis presented in Figure 5f shows a weak bimodal behaviour, with compositional clusters showing a phase with low relative Ni-content and one with high relative Ni content with slightly higher Al for the former. The nominal composition falls roughly between these two clusters suggesting that a low resolution average measurement of the composition would struggle to establish the presence of two distinct phases. This distribution is more similar to the parent compound in that strong variations in the relative Ni content are visible on the 10 – 100 nm scale.

For LNMA10, the chemical maps presented in Figure 5g show no evidence for a variation of the relative Ni content over a wide FOV, and the Al distribution appears to be similarly homogenized. The bivariate histogram presented in Figure 5h reveals that the bimodal distribution observed in LNMA01 has largely been replaced by an elongated ellipsoid whose center matches the nominal composition. These EDX maps do not show clear evidence for different compositions in the monoclinic and rhombohedral phases at the sampling resolution of these measurements.

These results indicate that the LMNO, LNMS10 and, to some extent, LNMA01 materials manifest themselves as two-phase systems on the 10 – 100 nm length scale. The Ni-rich R-phase either forms primarily at particle surfaces or, particularly in the case of Sn-doping, even segregates out as phase-specific particles. In contrast, LNMA10 appears to be more complex; Figure 5 indicates that it must either be a single-phase system, thereby contradicting the XRD, NPD, and NMR results, or that the individual phases intermix on the nanoscale in the form of nanodomains, and that the resolution at this FOV is insufficient to resolve them. Such nanodomain systems have been described in previous works[29], although it was noted that isolating these individual phases is very challenging in thick particles due to overlap in the beam projection direction.

## Thickness dependent nickel content

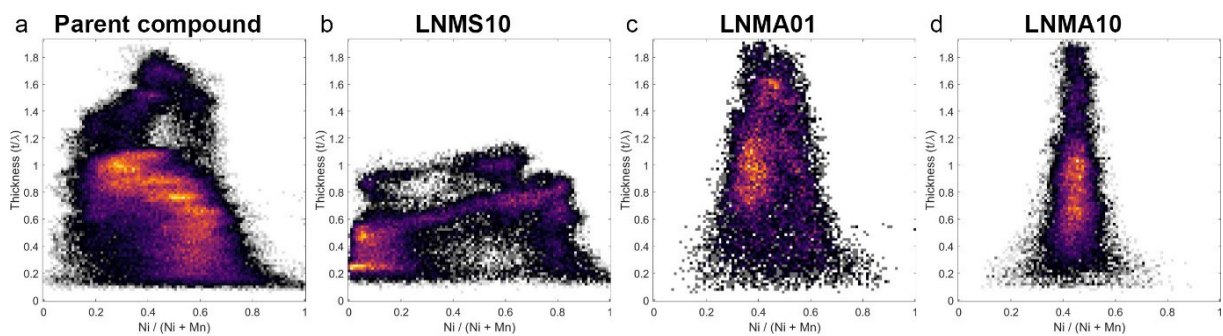

*Figure 6 - Bivariate histograms of thickness vs. relative Ni content for (a) LNMO, (b) LNMS10, (c) LNMA01, and (d) LNMA10.*

An alternative way to test for the presence of multiple phases with differing compositions at this wide FOV is to check how the relative Ni content varies as a function of the particle thickness. Each of the datasets in Figure 5 samples a wide range of thicknesses, which are calculated from the simultaneously acquired low-loss EELS datasets. As the thinnest regions correspond to the particle surfaces, if these surfaces are Ni-rich, one would expect to see an increase in the relative nickel content at the lowest thicknesses. Additionally, presenting the data in this manner also indirectly tests the domain size of the M- and R-phases, as each individual EDX spectrum is averaged over the particle thickness at that probe position. If a single particle consists of very small M- and R-phase domains, and if these domains are homogeneously distributed throughout the particles, thicker regions will converge on the average (nominal) composition, while thinner regions will diverge towards the composition of the individual phases as they are less likely to sample a random mixture of the two phases.

Figure 6 presents this analysis for the datasets shown in Figure 5 in the form of bivariate histograms revealing the relationship between the relative Ni content and the relative thickness as mean free paths ($t/\lambda$). LNMO (Figure 6a) shows a clear trend towards a Ni-rich phase at the thinnest regions, consistent with the observation that the edges are Ni-rich[24]. LNMS10 (Figure 6b) shows a clear separation between the phases in the thinnest regions, with few data points between the two clusters. Only at medium thicknesses around 0.6 $t/\lambda$ is significant mixing observed, representing areas where two individual particles with different composition spatially overlap. This behaviour is expected for a two-phase system with relatively large domain sizes. LNMA01 (Figure 6c) also shows a divergence from the nominal composition at lower thicknesses, with a slight preference for a Ni-rich phase at lower thicknesses, although this is more subtle than for LNMO. Finally, LNMA10 (Figure 6d) shows a relatively even divergence at lower thicknesses that would be consistent with the presence of very small nanodomains with no strict preference for accumulation of either phase at the particle surfaces.

### 2.6.2 High resolution STEM

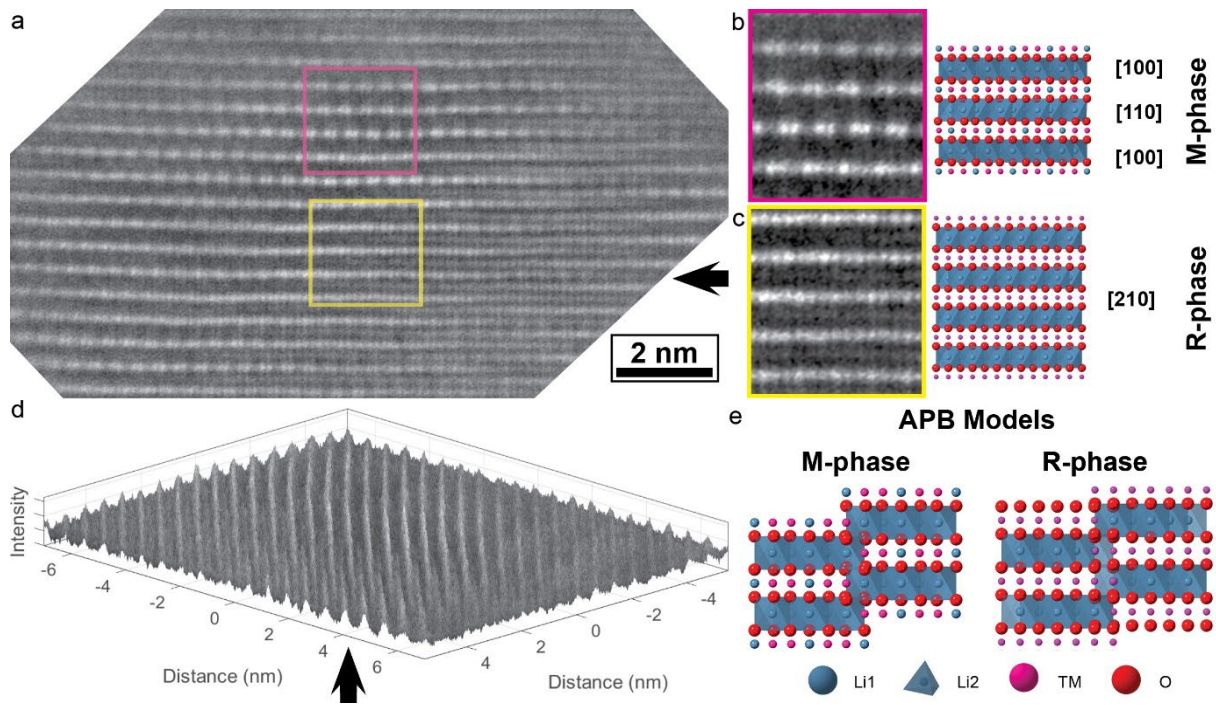

*Figure 7 – (a) HAADF images from a thin region of LNMA10 showing M-phase and R-phase nanodomains as well as antiphase boundaries. A mostly M-phase region is denoted by the magenta box and enlarged along with an atomic model in (b), while the same is done for the R-phase using a yellow box in (c). A 3D relief of (a) is provided in (d) in order to accentuate the presence of APBs as well as strain fields induced by a mismatch within the oxygen sublattice. The black arrows in (a) and (f) denote equivalent viewing directions. Atomic models for the proposed APB displacement for both M-phase and R-phase are provided in (e) along with a model legend.*

While the wide FOV results presented in Figure 5 and Figure 6 probe the chemical and morphological distribution of the M-phase and R-phase on the length scales of 10 – 1000 nm, higher resolution methods are needed to investigate the 0.1 – 10 nm length scales. Figure 7 summarizes the findings from LNMA10. In Figure 7a, a high resolution HAADF STEM from a thin region of the sample tilted is presented. The crystal was tilted to the R-phase [2 1 0] zone axis (equivalent to the M-phase [1 0 0] / [1 1 0] / [1 $\bar{1}$ 0] zone axes) to resolve the stacking of the 2D lithium-rich planes. Within this single image, two crystallographically distinct domains can be unambiguously identified, and these are marked and enlarged in Figure 7b (magenta) and Figure 7c (yellow). Since these are Z-contrast images, the heavier transition metal elements reveal the positions of the atomic columns and allow these individual domains to be indexed as the M-phase (Figure 7b) and R-phase (Figure 7c), respectively. Atomic models are provided next to the enlargements, and the stacking fault configuration characteristic of the M-phase is denoted. Thus, not only are both crystallographic domains in the composite LNMO structure identified, but it is also immediately evident that their individual domain size is approximately a few nanometers.

However, the existence of nanodomains is not the only feature observed in this sample, rather it is also evident that the 2D ordering of the transition metal and lithium-rich layers is inconsistent in multiple regions. This effect is best visualized by viewing parallel to the layers at a low elevation angle, and these conditions are simulated as a 3D relief, which is presented in Figure 7d. The black arrows in Figure 7a and d denote equivalent viewing directions. At present, our favoured explanation for this alternating stacking sequence is the presence of antiphase boundaries (APBs) having displacement vectors $\left[0 \ \frac{b}{6} \ \frac{c}{2}\right]$ in the M-phase and $\left[\frac{a}{3} \ \frac{b}{3} \ \frac{c}{6}\right]$ in the R-phase. Similar defect structures have been observed in Co-containing Li(Ni$_{1-x-y}$Co$_x$Mn$_y$)O$_2$[30], pure Li$_2$MnO$_3$[31], as well as in an all solid-state spinel LiNi$_{0.5}$Mn$_{1.5}$O$_4$[32]. There, it was argued on theoretical grounds that the addition of dopants such as Al should reduce the number of APBs, although concentrations as high as LNMA10 were not investigated. Atomic models for the proposed defect structure in our system are provided in Figure 7e. While these models are restricted to M-phase [1 0 0] and R-phase [2 1 0] domains, it is apparent that the equivalency extends to M-phase [1 1 0] and [1 $\bar{1}$ 0] orientations as well as M-phase / R-phase boundaries.

This observation suggests that such APBs can form provided that the atomic positions of cations in the 2D transition metal and lithium-rich layers can be nearly exactly matched, which appears to be the case for this dopant concentration. However, the congruency of the cation sublattice does not extend to the oxygen sublattice, which instead exhibits a slight mismatch. This results in local strain, which is also best visualized in Figure 7d. This perspective allows us to easily identify a slight buckling of the TM-rich layers as they extend away from the area of APB overlap, as well as a rotation to the TM layers, which are both likely strain-induced effects. Thus, we observe that the nanoscale microstructure of LNMA10 exhibits a relatively large number of complex phase boundaries and local strain fields that arise from the formation of nanodomains.

Similar high resolution STEM investigations were carried out on LNMA01, and these are summarized in Figure 8. As in Figure 7, a thin crystal was tilted to the same zone axis and the edge of this crystal was studied. Figure 8 presents both an HAADF (a) and iDPC (b) overview of this region. iDPC is a technique that maps the center-of-mass (COM) of the transmitted electron probe[33] and, as such, is a linear contrast mechanism that can be used to simultaneously image both light and heavy elements in well-oriented crystals, including Li in layered Mn oxide batteries[34]. Similar to LNMA10, both the M-phase and R-phase are readily identifiable in this image. However, in contrast, these phases are not present as nanodomains but, rather, are morphologically confined to the surface (R-phase) and bulk (M-phase). Additionally, a thin spinel phase is observed at the very surface. These conclusions are drawn via inspection of three distinct local Fourier transforms taken from three ROIs denoted in Figure 8b, with the transforms themselves appearing in Figure 8c-e along with the crystal structure and orientation that most closely matches the observed spatial frequencies. Figure 8c indexes most closely to a spinel-like phase (S-phase) in the [1 1 0] orientation, and this is observed closest to the edge of this crystal. Such spinels have been observed many times before at crystal surfaces, and are noted to be very thin and, thus, difficult to resolve using XRD. The cyan box located a few nanometers away from the surface returns the Fourier transform observed in Figure 8d. This indexes very closely to the R-phase in the [2 1 0] orientation and is estimated to extend from approximately 1 nm away from the surface to 6 nm into the bulk. Finally, the Fourier transform from magenta box presented in Figure 8e closely matches the M-phase in the [1 0 0] / [1 1 0] / [1 $\bar{1}$ 0] orientation, with the characteristic stacking fault streaks. Crucially, unlike LNMA10, this crystal reveals a largely unobstructed Li-ion diffusion pathway from the surface into the bulk with little to no stress or buckling observed.

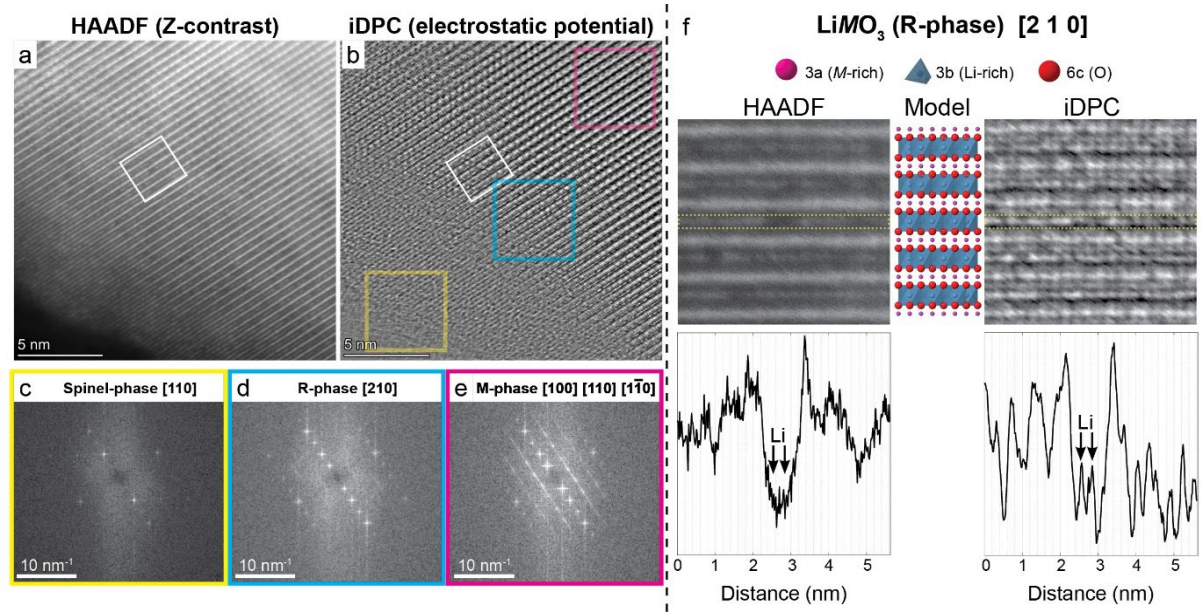

*Figure 8 – Simultaneously recorded HAADF (a) and iDPC (b) STEM images from a thin crystalline region in LNMA01. In (b), three boxes are presented from which Fourier transforms have been computed in (c – e). A white box in (a) and (b) denotes the enlarged region presented in (f) after image rotation. A dotted yellow line in (f) denotes the integration area for the line scans below. The atomic model in (f) is constructed from the R-phase (Li$_2$MO$_3$) in the [2 1 0] orientation and the Wyckoff positions are labelled above. Peaks observed in the line scan of the iDPC image that are absent from the HAADF image are interpreted as arising from pure Li scattering centers on the 3b site, while neighbouring atomic sites in this layer exhibit Li / TM mixing.*

Close inspection of the R-phase region in Figure 8a reveals additional information about the site-specific cation population in this crystal. Since the contrast mechanism is primarily sensitive to the atomic mass, this image is particularly sensitive to the location of the transition metals. Nearly all of the 3a sites observed in this crystal return similar intensities, thus suggesting that they are uniformly populated with transition metals. However, the observed intensities on the 3b sites varies considerably in a seemingly random manner. The white box in Figure 8a and b picks out a region from the HAADF and iDPC images that illustrates this effect. This selection is rotated and enlarged in Figure 8f. Additionally, line profiles that were vertically integrated over one of the R-phase (0 0 6) planes that captures the 3b sites (denoted by the dotted yellow lines) is presented below each image. This particular plane was chosen to highlight the contrast difference between the two imaging modes.

The line profile from the HAADF image in Figure 8f shows a clear loss of contrast in the dark region of this atomic plane. This would be expected if these particular atomic sites were to be primarily populated with lithium atoms, as their low scattering power renders them largely invisible in HAADF imaging. However, the same region in the iDPC image reveals a clear contrast that can be directly ascribed to the 3b sites, albeit with a much lower intensity than the neighbouring sites in this plane. As iDPC is more sensitive to lighter elements, this contrast is therefore best explained by the presence of atomically-localized weak scattering centers in the form of columns of lithium atoms on this site. Consequently, by combining the information from both the HAADF and iDPC images, we argue that these images directly show that the Li-rich 3b site in the R-phase is randomly populated with transition metals and/or Al, which has been hypothesized to stabilize the host system during delithiation[17,19].

## 3. Discussion

The electrochemistry results clearly show that the performance of these Co-free LIB cathodes is strongly influenced by chemical doping with Al and Sn. $Sn^{4+}$ doping reduces the discharge capacity in LNMS01 and LNMS10 from 140 $mAhg^{-1}$ to 60 $mAhg^{-1}$, respectively (Figure 1), in agreement with reported results[21,22]. In contrast, small amounts of Al increase the discharge capacity and cycling stability, even when compared to the undoped parent compound. These effects are difficult to explain by assuming a solid solution of dopant material, requiring a more rigorous structural, chemical, and nanoscale investigation to understand, as outlined in this manuscript.

The XRD results summarized in tables 2 and 3 as well as in Figure 3 reveal that all of the samples investigated here, regardless of dopant type or level, are structurally best described as a two-phase system consisting of both the R-phase and M-phase. Sn doping leads to a strong divergence in unit cell parameters for the two phases (Figure 3), while phase-specific compositions derived from NPD refinement (tables 2 and 3) support the interpretation that Sn and Ni both preferentially migrate into the R-phase, leading to a Mn-rich M-phase. We interpret this to mean that the introduction of Sn leads to decreasing structural and chemical compatibility between the two phases, forcing them to separate into individual phase-pure nanoparticles with minimal mixing.

In contrast, the introduction of Al to the parent structure causes a convergence of the unit cell parameters for both crystallographic phases (Figure 3), while the Ni to Mn ratio remains closer to the parent compound. The origin for this convergence appears to be that Al is observed to enter both phases, resulting in a preferential migration of Ni from the R-phase to the M-phase in an effort to maintain charge balance. Consequently, the unit cell volumes of both phases converge to similar values, increasing their structural and chemical compatibility. This manifests itself as an increasingly intimate domain mixture on the nanoscale as the amount of Al doping increases.

This interpretation of a decrease in the domain size of the M- and R-phases is first provided from the NMR study (Figure 4). The lack of significant broadening due to BMS effects in LNMS10 strongly suggests that, even for the highest dopant concentration, both the R- and M-phases are structurally and chemically distinct and have large domain sizes. Contrarily, the severe broadening observed for the LNMA10 sample is best explained by assuming that both phases are intimately mixed and present in the form of nano-sized domains. Crucially, this broadening is also observed for the LNMA01 sample, suggesting that some degree of mixing has taken place already at these low dopant concentrations.

The spatial distribution of these two phases along with their composition is explored through the TEM results at two different length scales: tens to hundreds of nanometers (Figure 5 and Figure 6) and sub-nanometer (Figure 7 and Figure 8). The wide FOV EDX maps and bivariate histograms from Figure 5 reveal that the interparticle phase segregation observed in the parent compound becomes exacerbated to full individual particles when doping with Sn. Combined with the other results, this indicates that Sn doping leads to two distinct phases that fail to intermix on the nanoscale. As phase intermixing is known to be a prerequisite for higher voltage retention, we conclude that this is the proximate cause for the degraded electrochemical performance of these samples observed in Figure 1. Conversely, the introduction of Al causes a much more intimate phase mixture, such that it is no longer even possible to confirm the presence of two chemically-distinct phases in the LNMA10 sample at the resolution presented in Figure 5. However, the thickness-dependent relative nickel content presented in Figure 6 does suggest that some sort of domain mixing is occurring, albeit at length scales that are shorter than the particle thickness, which we refer to as nanodomains.

Direct visualization of these nanodomains as well as their influence on the lithium diffusion pathways requires both a sufficiently thin sample as well as sub-nanometer spatial resolution imaging, which we present in Figure 7 and Figure 8 for LNMA10 and LNMA01, respectively. High resolution HAADF imaging of the LNMA10 sample (Figure 7) provides structural evidence for the presence of not only M-phase and R-phases nanodomains with lateral dimensions on the order of a few nanometers, but also antiphase boundaries. The presence of nanodomains along with antiphase boundaries disrupts the coherency of the oxygen sublattice, resulting in localized strain fields and disordering which we believe conspire with the larger number of interfaces to hinder ionic transport in this system, consistent with the degraded electrochemical properties from Figure 1 as well as the reduced lithium diffusivity from Figure 2. For the LNMA01 sample, these nanodomains appear to be morphologically segregated to the particle surface (R-phase) and particle bulk (M-phase), with an additional thin S-phase observed at the vacuum / particle boundary. In contrast to LNMA10, the M-phase and R-phase boundary is highly coherent with no disruption of either the oxygen sublattice or the lithium ion pathways (save for at the thin S-phase surface). High resolution STEM imaging combined with DPC (Figure 8) moreover confirms that the 3b-site in the R-phase is randomly populated with approximately a small percentage of transition metal and/or Al atoms. This thus constitutes a direct atomic-scale observation of structural stabilization in this system, consistent with the inferred findings from previous studies in Co-containing compounds[17,19] as well as theoretical works on Co-free systems[20].

## 4. Conclusions

In this manuscript, we outline how doping $Li_{1.1}Ni_{0.35}Mn_{0.55}O_2$ with small amounts of Al or Sn influences the size, composition, and distribution of the M- and R-phases on the nanoscale. Through combined use of NPD, aberration-corrected STEM, EDX, EELS, and NMR measurements, we are able to deduce that the composition, lateral dimensions, and spatial distribution of these domains varies considerably depending on the type and amount of dopant used. Doping with Sn results in a strong phase segregation, while doping with Al homogenizes the system both compositionally and structurally. A small amount of Al leads to a structural improvement to the lithium diffusion pathways, while too much Al results in overmixing characterized by the formation of nanodomains and the subsequent disruption of these pathways. We therefore conclude that this study provides researchers with unique insight into how chemical doping can be parameterized to control the spatial distribution and local chemistry of the M-phase and R-phase in this system, yielding a tool with which the electrochemical performance can be optimized in this and other layered manganese oxide systems.

## 5. Experimental Section

### 5.1 Sample series fabrication

All samples were fabricated using the spray pyrolysis process. An aqueous solution was prepared by dissolving stoichiometric amounts of metal nitrates in distilled water mixing individual precursor solutions under stirring. The homogeneous solution was atomized by a two-phase nozzle (pressurized air + solution) into a pre-heated rotating (~2rpm) furnace (Entech Energiteknik AB) under constant air flow, yielding an approximate average residence time of ~2 seconds at 900 °C. This caused instant water vaporization, and onset of nitrate decomposition, as mixed metal oxide began to form. The collected powders were calcined using a Nabertherm NW300 chamber furnace at 900 °C for 6 h in air with heating and cooling rates of 200 °C per hour, as this was determined to be the optimal temperature for phase segregation in a previous study[24].

### 5.2 Electrochemistry tests

The electrodes for electrochemical testing were prepared by mixing 85 wt% active material with 10 wt% conductive carbon Super-P (Alfa Aesar) and 5 wt% polyvinylidene fluoride (PVDF, Sigma-Aldrich) dissolved N-methyl pyrrolidinone (NMP) by ball milling (Retsch® MM400) for 45 min at 15

Hz with 16.67 g of stainless steel balls so the active material/balls weight ratio was 0.12. The obtained slurry was coated on Al current collector using a laboratory doctor blade and dried at 65 °C overnight. The dried foils were punched out to be circular electrodes (12 mm diameter) and again dried under vacuum at 100 °C for 24 h before being transferred into an Ar-filled glovebox ($H_2O$ < 0.5 ppm). The active material loading was 1.5 mg cm$^{-2}$. The electrochemical performance of all the samples was tested in half cell coin cells (2032R). A lithium metal foil pressed into a stainless steel current collector disc was used as the negative electrode, and a microporous monolayer polypropylene membrane (Celgard 2400) was used as a separator. The separator was soaked with electrolyte solution 1.2M $LiPF_6$ in ethylene carbonate (EC)/ethyl methyl carbonate (EMC)/dimethyl carbonate (DMC) (2:5:3 volume %) + 2 wt% vinylene carbonate (VC) + 2 wt% fluoroethylene carbonate (FEC) supplied by Solvionic. After an overnight open circuit potential period, the cells were tested at constant current in the potential window of 4.8 and 2 V vs $Li^+$/Li using a VMP3Z biologic multichannel instrument at room temperature. For GITT, the cells were subjected to constant current pulses of C/20 for 30 minutes followed by a relaxation time interval of 2 h.

### 5.3 X-Ray Diffraction

X-ray powder diffraction (XRPD) patterns were collected using Cu-Kα radiation and a Brukker D2 phaser table-top diffractometer. Neutron powder diffraction (NPD) data were collected at the ISIS time-of-flight neutron source at the Rutherford-Appleton laboratory in UK with the POLARIS diffractometer and samples placed in 6 to 8 mm diameter V cans. Structure refinements by the Rietveld method were carried out with the FullProf program [P1] and using NPD data from the 147° and 52° banks.

### 5.4 Nuclear magnetic resonance

$^7$Li MAS NMR experiments were performed at the magnetic field strength of 14.1 T (233.23 MHz Larmor frequency) with a Bruker Avance-III NMR spectrometer equipped with a 1.3 mm MAS probehead. A 60.00 kHz MAS rate was employed. Acquisitions involved a rotor-synchronized, double-adiabatic spin-echo sequence with a 90° excitation pulse of 1.00 µs followed by a pair of 50.0 µs tanh/tan short, high-power adiabatic pulses (SHAPs) with 5 MHz frequency sweep[35,36]. All pulses operated at the nutation frequency of 250 kHz. 4096 signal transients with 0.5 s relaxation delay were collected. The $^7$Li chemical shifts were referenced with respect to solid LiF.

### 5.5 Transmission electron microscopy

TEM experiments were performed on a double aberration-corrected Themis Z instrument (Thermo Fisher) operated at 300 kV. Three samples were chosen for TEM investigation: Sn x = 0.10, Al x = 0.01, and Al x = 0.10. These samples were prepared for TEM analysis by diluting a small amount of powder in isopropanol alcohol, ultrasonicating it for approximately 30 minutes, and dispersing a few drops onto a thin carbon grid. STEM experiments were performed by correcting pre-specimen aberrations up to 5$^{th}$ order with a CEOS DCOR corrector and then scanning a finely focused electron probe across the thin specimen. EELS and EDX experiments were performed simultaneously using a probe current varying between 100 – 250 pA for each experiment, depending on the field-of-view (FOV). EELS was collected using a post-column Gatan Image Filter (GIF, Gatan Inc.) using a convergence angle of 21.4 mrad and a collection angle of 23 mrad. iDPC images were acquired on a 4-quadrant annular dark field detector (Thermo Fisher) using a probe current of 50 pA.

# Acknowledgements

This research was funded by HORIZON 2020-supported EU project COBRA, grant number H2020-EU.3.4.-875568. The authors also acknowledge Stephen hull and Ron Smith for collecting neutron data at the ISIS neutron and muon source using rapid access. The experiment number was 2000189.

# Conflict of Interest

The authors state no conflict of interest.

# Data Availability Statement

The data and analysis methods supporting the results presented in this study are available upon request from the corresponding author.